\documentclass[11pt,aps,prc,superscriptaddress,showpacs,floatfix]{revtex4-1}
\usepackage[dvipdfmx]{graphicx} 
\usepackage{epsfig}
\usepackage{bm}
\usepackage{amssymb}
\usepackage{wrapfig}
\usepackage{ulem}
\usepackage{multirow}
\usepackage{color}

\newcommand{\psla}{\ooalign{\hfil/\hfil\crcr{p}}}
\newcommand{\ksla}{\ooalign{\hfil/\hfil\crcr{k}}}

\newcommand{\epsisla}{\ooalign{\hfil/\hfil\crcr{$\epsilon$}}}

\newcommand{\el}{{\cal L}}

\newcommand{\psibar}{\mbox{$\overline{\psi}$}}

\newcommand{\Ubar}{\mbox{$\overline{U}$}}

\newcommand{\tldu}{\mbox{${\tilde u}$}}

\newcommand{\vp}{\mbox{$\bm{p}$}}
\newcommand{\vq}{\mbox{$\bm{q}$}}
\newcommand{\vbr}{\mbox{$\bm{r}$}}
\newcommand{\vk}{\mbox{$\bm{k}$}}

\newcommand{\vA}{\mbox{$\bm{A}$}}

\newcommand{\vQ}{\mbox{$\bm{Q}$}}

\newcommand{\vepsi}{\mbox{\boldmath $\epsilon$}}

\newcommand{\br}{\bf{r}}
\newcommand{\bp}{\bf{p}}
\newcommand{\bq}{\bf{q}}

\newcommand{\bk}{\bf{k}}

\begin{document}

\title{Compton Scattering of $\gamma$-Ray Vortex 
with Laguerre Gaussian Wave Function}

\author{Tomoyuki~Maruyama}
\affiliation{College of Bioresource Sciences, Nihon University,
Fujisawa 252-8510, Japan }
\affiliation{National Astronomical Observatory of Japan, 2-21-1 Osawa, Mitaka, Tokyo 181-8588, Japan}
\author{Takehito~Hayakawa}
\affiliation{National Institute for Quantum and Radiological Science and Technology, Tokai, Ibaraki 319,-1106, Japan}
\affiliation{National Astronomical Observatory of Japan, 2-21-1 Osawa, Mitaka, Tokyo 181-8588, Japan}
\author{Toshitaka~Kajino}
\affiliation{Beihang University, School of Physics and 
Nuclear Energy Engineering,\\
Int. Center for Big-Bang Cosmology and Element Genesis, Beijing 100191,
China}
\affiliation{National Astronomical Observatory of Japan, 2-21-1 Osawa, Mitaka, Tokyo 181-8588, Japan}
\affiliation{The University of Tokyo, Bunkyo-ku, Tokyo 113-0033, Japan}
%

%\pacs{12.60.Fz, 32.80.Cy}

\begin{abstract}
In this work, we report calculation for Compton scattering of 
a $\gamma$-ray vortex
with Laguerre Gaussian wave function on an electron in the framework of
the relativistic quantum mechanics.
We have found the following unexpected feature.
The momentum of scattered photon distributes outside of the reaction
plane determined by the incident photon and the scattered electron, 
and hence the energy of the scattered photon also distributes.
This novel result indicates that one can identify a $\gamma$-ray vortex by
measuring coincidentally the scattered angles of the electron and photon. 
 \end{abstract}

\maketitle

%%%%%%%%% Introduction
Photon vortices carrying orbital angular momentum  \cite{Allen92} are 
interesting both from the fundamental research \cite{BPA94, ACD06, PBMVPCR10}
 and for applications  \cite{Yao11, CWSJZ12, Wang12, LSBP13,
TTTTMMO13, Afanasev13, BYRTKHWR13}. 
Furthermore, it is suggested that the photon vortex can  be created in astronomical system \cite{TTMA11}.
Recently, it has been proposed 
to generate $\gamma$-ray vortices in the MeV energy region using laser
Compton scattering with laser vortex
\cite{Jentschura11a,Jentschura11b,Petrillo16} and nonlinear laser
Compton scattering with highly intense circularly polarized laser
\cite{Taira17}.
When $\gamma$-ray vortices are  available in laboratory, 
they open a new frontier in nuclear and particle physics \cite{Taira17,Ivanov11}.
However, there is no practical method to identify $\gamma$-ray 
vortices in the MeV energy region. 
Compton scattering is a dominant process of photons and atoms
in the energy region from several hundred keV to several MeV. 
It is well known that the differential cross-section of Compton
scattering of linearly polarized $\gamma$-rays depends on the angle
between the scattering plane and the polarization plane. 
Thus, polarimeters based on Compton scattering have been used in
nuclear physics \cite{Jones95} and $\gamma$-ray astronomy \cite{Lei97}. 
%Thus, it is important to develop a method to identify $\gamma$-ray
Thus, it is important to calculate the cross section of Compton scattering with $\gamma$-ray
vortices.

\begin{figure}[htb]
\begin{center}
\vspace{-1em}
{\includegraphics[scale=0.55]{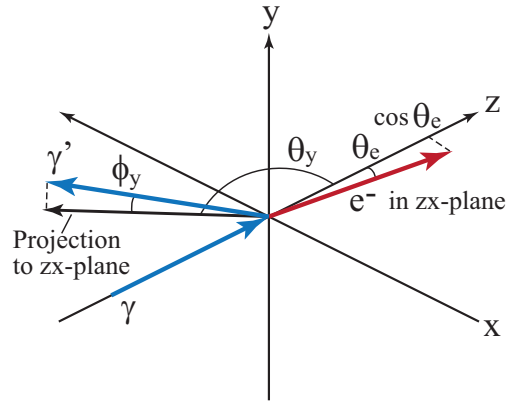}}
\caption{\small
Coordinate system used in the calculation. $\gamma$ and $\gamma$' denote the initial and final photons, respectively. The electron as e$^-$ is scattered in the $zx$-plane.
}
\label{coordinate}
\end{center}
%\end{figure}
\end{figure}

%%%%% Calcualtion
We consider Compton scattering of a $\gamma$-ray vortex with a wave
function of Laguerre Gaussian (LG) on an electron at rest.
We also consider coincidence measurements of the scattered electron and photon. 
We set the system so that an initial photon propagates along 
the $z$-direction with the energy $k$ 
and the electron is scattered in the $zx$-plane 
(see Fig.~\ref{coordinate}).
% We also define $y$-axis as the direction perpendicular to the $zx$-plane.
%%%
%%%%%%%%  Comp]ton scattering in elativistic quantum mechanics 
The amplitude of Compton scattering
in relativistic quantum mechanics \cite{Bjorken} is given by
\begin{equation}
 S_{if} = e^2 \int d^4 x d^4 y \psibar_f (x) \gamma^\mu S_F (x,y) \gamma^\nu 
\psi_i(y) \left[ A_f^{\mu *} (x) A_i^\nu (y) 
+  A_i^\mu (x) A_f^{\nu *} (y) \right] ,
\end{equation}
where $e$ is the elementary charge, 
$\psi_i$ and $\psi_f$ are the initial and final electron wave
functions, respectively, $S_F$ is the electron propagator, and 
$A_i \equiv (A_i^0, \vA_i)$ and $A_f \equiv (A_f^0, \vA_f)$ are 
the initial and final photon fields, respectively.
Though the momenta of photon vortices are under off-mass-shell condition,
it is expected that the on-mass-shell condition is satisfied 
for observed photons.
Therefore, we assume that the final photon wave function is 
the plane wave with 
the final photon momentum of $q \equiv (|\vq|, \vq) = (|\vq|, \vq_T, q_z$).

We choose the Lorentz gauge and  $A_0 =0$ for the photon field and 
write the electron and photon fields as
\begin{eqnarray}
&& \psi(x) = \frac{1}{\sqrt{\Omega}}U(\vp, s) e^{i\bp \br - i E_p t} ,
~ %\nonumber \\ && 
\vA_i(\vbr) =  \frac{\vepsi_i (h_i)}{\sqrt{2 k}} 
 u (\vbr) e^{ - i k t} ,
~
\vA_f (\vbr) =  \frac{\vepsi_f (h_f)}{\sqrt{2 |\vq| \Omega}}
 e^{i \bq \br - i |\bq| t} ,
%%s
\end{eqnarray}
where 
$\Omega$ is the volume of the system, $U(\vp,s$ is the Dirac spinor 
of an electron with the momentum $\vp$ and the spin $s$, 
$k$ is the energy of the initial photon, 
$h_{i (f)}$ indicates the helicity of the initial (final) photon,
and the polarization vector satisfies
$\vepsi_f \cdot \vq =0$ and $\vepsi_f \cdot \vepsi_f = 1$.
We write  the initial and final momenta of the electron as 
$p_i = (E_i, \vp_i)$ and $p_f = (E_f, \vp_f)$, respectively.
The scattering amplitude is rewritten as 
\begin{eqnarray}
S_{if} & =& \frac{e^2}{2 \sqrt{k_i^0  |\vq|  \Omega}}
\Ubar (\vp_f,s_f) \left[ \epsisla_f S_F (p_f + q) \epsisla_i
 +  \epsisla_i  S_F (p_i - q)  \epsisla_f
\right] U(\vp_i,s_i)
\nonumber \\ &&  \qquad\qquad\qquad \times
 \tldu (\vp_f + \vq - \vp_i)  (2 \pi) \delta (E_f + |\vq| - E_i - k) 
\end{eqnarray}
with $\epsilon_{i,f} = (0, \vepsi_{i,f})$ and 
\begin{eqnarray}
&& S_F(p) = \frac{\psla +m}{p^2 +m^2 + i \delta} , \quad
\tldu (\vk) = \int d \vbr e^{-i \bk \cdot \br} u(\vbr).
\end{eqnarray}
Then, the cross-section is given  by
\begin{eqnarray}
&& d \sigma =
\frac{e^4}{4 k E_i} 
 W_{if} \left| \tldu (\vp_f + \vq - \vp_i) \right|^2
(2 \pi) \delta (E_f + |\vq| - E_i - k )
\frac{d \vq}{(2 \pi)^3 |\vq|}
\frac{d \vp_f}{(2 \pi)^3 E_f}
\end{eqnarray}
with
\begin{eqnarray}
W_{if}&=& E_i E_f \left| \Ubar (\vp_f,s_f) 
\left[ \epsisla_f (h_k)  S_F (p_f + q) \epsisla_i + 
\epsisla_i S_F (p_i - q)  \epsisla_f(h_k) \right] U(\vp_i, s_i) \right|^2 .
\end{eqnarray} 

We assume that the initial electron is at rest, that the initial photon
is parallel to $z$-direction, and 
that the final photon polarization is not observed.
Then, we substitute $p_i = (m, 0, 0, 0)$ and 
$\vepsi_i (h_i)= (1, i h_i, 0)/\sqrt{2}$ with $h_i = \pm 1$.
We average the spin of the initial electron and sum over 
the spin of the final electron and the polarization of the final photon.
In addition, we rewrite $p_f = (E_p, \vp_e) = (E_p. \vp_T, p_z)$.
As the result, the cross-section is written as
\begin{eqnarray}
&& d \sigma = \frac{\alpha^2 w_0^2}{8 \pi^3 m k |\vq| E_p}
{\bar W}_{if} |\tldu(\vp + \vq)|^2  \delta  (E_p + |\vq| - m -k)
d \vq d \vp ,
\label{dCrs}
\end{eqnarray}
with
\begin{eqnarray}
&& {\bar W}_{if} = \frac{1}{2} \sum_{s_i, s_f, h_f} W_{if}
%%
%%%
\nonumber \\ &=& \frac{1}{8} \sum_{h_k} {\rm Tr} 
\left\{ (\psla_f +m)  \left[ \frac{\left( 2p_f \cdot \epsilon_f (h_k)
-  \ksla_f \epsisla_f (h_k) \right)  \epsisla_i}
{2p_f \cdot k_f} 
- \frac{\epsisla_i \left(\epsisla_f (h_k) 
\ksla_f + 2 p_i \cdot  \epsilon_f (h_k) \right)}
{2p_i \cdot  k_f }  \right] 
\right. \nonumber  \\ && \left. \qquad\qquad \times
 (\psla_i +m) \left[ 
\frac{  \epsisla_i^* \left( 2p_f \cdot \epsilon_f -  \epsisla_f \ksla_f \right)}
{2p_f \cdot k_f} 
- \frac{ \left(\ksla_f \epsisla_f + 2 p_i \cdot \epsilon_f \right) \epsisla_i^*  }
{2p_i \cdot  k_f } \right]
  \right\} 
%%%%%%
\nonumber \\ &=&
 \frac{1}{2} \left\{
\frac{m q_z^2}{|\vq| (p_f \cdot q)}
+ \frac{m k}{(p_f \cdot q)^2}
\left[ |\vp|^2 - \frac{(\vp \cdot \vq)^2}{|\vq|^2} \right]
\right. \nonumber \\ && \left. \qquad\qquad
+ \frac{E_p |\vq| - p_z q_z}{m |\vq|} 
+ \frac{p_{z}}{(p_f \cdot q)} 
\left[ \frac{ q_z (\vp \cdot \vq) }{|\vq|^2} - p_z 
 \right] \right\}.
%%%
\label{AvWel}
\end{eqnarray}
Note that these cross-sections are independent of the initial photon
helicity $h_i$.
%

%%%%%%%%% Laguerre-Gaussian wave photon
In this work, we consider the LG wave  \cite{Allen92} for the initial photon
which is written as
\begin{eqnarray}
&& u (\vbr) = \sqrt{\frac{2}{\pi R_z}} 
 \frac{1}{w(z)} G \left[ |L|, p, \frac{r}{w (z)} \right]
%%%%
\exp \left\{ i \left[ L \phi + k z +  \frac{z r^2}{z_R w^2(z)}
% + \frac{k r^2}{2R(z)} 
 - (2p + |L| + 1) \theta_G \right] \right\} 
\label{LGr}
\end{eqnarray}
with
\begin{eqnarray}
&&  G \left[ |L|, p,  x \right] = \sqrt{\frac{p!}{\pi(|L| + p)!}}
\left( \frac{x}{\sqrt{2}} \right)^{|L|}  e^{-x^2/4} 
\el^{|L|}_p \left( \frac{x^2}{2} \right) ,
\nonumber \\
&& \theta_G = \tan^{-1} \left(\frac{z}{z_R}\right), \quad
w (z) = w_0 \sqrt{1 + z^2/z_R^2}, \quad
% R(z) = (z^2 + z_R^2)/z, 
\quad z_R =  k w_0^2 /2 .
\label{Const}
\end{eqnarray}
where $L$ is an integer indicating the orbital angular momentum 
of the initial photon, 
$\el_p^{|L|}$ is the associated Laguerre function, $R_z$ is the size
of the system along the $z$-direction, and $w_0$ is the waist radius at $z=0$.
%
%%%%% Fourier transformation 
The Fourier transformation of $u(\vbr)$ becomes
\begin{eqnarray} 
&& \tldu(\vQ)
= \sqrt{\frac{(2 \pi)^3}{R_z} }
e^{i (p + |L|/2 ) \pi}e^{i L \phi_q }
 w_0  G \left[ |L|, p ; w_0 Q_T \right] 
\delta \left( Q_z + \frac{Q_T^2}{2k} - k \right)
\label{PhAmp}
\end{eqnarray}
with $Q_T \equiv \sqrt{Q_x^2 + Q_y^2}$ and
$\phi_q$ being the azimuthal angle of the momentum $\vQ$ 
along $z$-axis.
Here, we consider coincident measurements of the scattered photon and
electron. 
%It is well known that,  in the case of the initial photon with plain wave, 
%the final photon momentum is on $zx$-plane.
%
When the initial photon is the plane wave with momentum $(0,0,k)$,
the final electron momentum is  
$\vp_e / p_e = (- \sin \theta_e, 0, \cos \theta_e)$ with
$p_e = 2m k (k+m) \cos \theta_e/[(k+m)^2 - k^2 \cos^2 \theta_e]$ and 
the final photon momentum is
$\vq = \vq_0 = |\vq_0| ( \sin \phi_0, 0, \cos \phi_0 ) 
= (p_e \sin \theta_e , 0, k - p_e \cos \theta_e)$.
In contrast, when the incident photon is the LG wave, 
the final photon momentum distributes, while the initial photon momentum
is given by $\vQ = \vp_e + \vq$,  thereby satisfying the relation
$p_{z}+q_z + (\vp_{T} + \vq_T)^2/2k - k = 0$ [see Eq.~(\ref{PhAmp})].
To illustrate the final photon momentum distribution, we take
$y$-direction to be a new principal axis and write the final photon
momentum as 
$ \vq = |\vq| 
(\cos \theta_y \sin \phi_y, \sin \theta_y, \cos \theta_y \cos \phi_y)$
as shown in Fig.~\ref{coordinate}.
%
%%%% Cross section with coincidence measurement at the fixed electron momentum
Combining Eqs.~(\ref{dCrs}) and (\ref{PhAmp}), 
we obtain the cross-section for the incident photon of LG wave.
By integrating  the cross-section over $|\vq|$ and $\phi_y$ 
for a fixed electron momentum, the cross-section is written as 
\begin{equation}
\frac{d^4 \sigma}{d \vp_e^3 d \sin \theta_y}
= \frac{\alpha^2 w_0^2 |\vq|}{4 \pi m E_p 
\left|(k-q_z)q_x - q_z |\vp_T| \right|}
{\bar W}_{if} \left[ G (L, p; w_0|\vp_T + \vq_T)|) \right]^2  
%%%%%%%%
\end{equation}
with $|\vq| = k + m - E_p$.
An important consideration in measurements is the  divergence angle
of the incident photon, 
which is determined by the photon energy $k$ and the waist radius $w_0$.
We take the incident photon energy to be $k = 500$~keV, where 
Compton scattering dominates.
Although $w_0$ is a free parameter in the present calculation, 
it is determined by the generation mechanism of $\gamma$-ray vortices 
in the framework of the quantum mechanics.
However, to our knowledge, there is no theoretical prediction.
Thus, we take $w_0$ to be 25~pm that is approximately ten times 
of the wave length of the present incident photon (2.48~pm)
in the following discussion. 
%

\begin{figure}[htb]
\begin{center}
%\vspace{-1.5em}
{\includegraphics[scale=0.7]{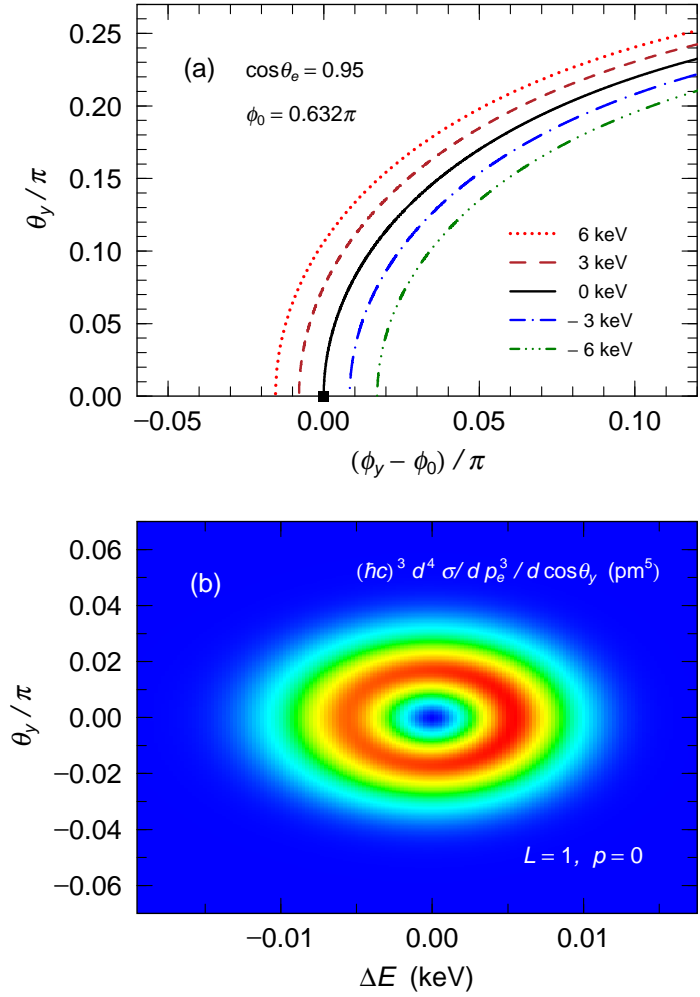}}
\caption{\small
Upper panel (a): the directions of the scattered photons at 
the fixed  momenta of the final electrons, when $\cos \theta_e = 0.95$.
The dotted, dashed, solid, dot-dashed, and dpt-dot-dashed lines
represent the results when $\Delta E = 6$~keV, 3~keV, 0~keV, $-3$~keV,
and $-6$~keV, respectively.
Lower panel (b): the contour plots of the differential cross-section of 
Compton scattering 
integrated over the azimuthal angle $\phi_y$
when  $L=1$, $p = 0$, and $\cos \theta_e = 0.95$.
The horizontal axis shows the energy difference $\Delta E$, and
the vertical axis shows the polar angle along the $y$-axis.
}
\label{dSgEA}
\end{center}
\end{figure}

%%% Result and Discussion
We find two unexpected features.
First, the differential cross-sections have finite values out of
the $zx$-plane.
Figure \ref{dSgEA}(a) shows the angles of the scattered photons which give the finite cross-sections justifing the relation
$p_{z}+q_z + (\vp_{T} + \vq_T)^2/2k - k = 0$
when $\cos \theta_e = 0.95$.
The results exhibit that the distributions of the cross-sections are 
perpendicular to the horizontal axis at $\theta_y = 0$, 
clearly implying that the strengths of the cross-sections 
distribute out of the $zx$-plane. 
Second, the energy of the scattered photon is shifted from
that of the standard Compton scattering.
In addition, the energy shift correlates with the spatial shift from the $zx$-plane.
The lower panel [Fig. \ref{dSgEA}(b)] shows the cross-sections for the LG waves with $L=1$
and $p=0$ as a function of the energy shift  
$\Delta E = |\vq| - |\vq_0|$ and the scattered angle to $y$-axis,
$\theta_y$.
It should be emphasized that the strengths are exactly zero at $\phi_y = 0$ and $\Delta E = 0$, at which the strength appears in the standard Compton scattering.
This corresponds to the fact that the amplitude of the LG wave function is 
zero along the beam line when $L \ne 0$.

\begin{figure}[htb]
\begin{center}
\vspace{-1.5em}
{\includegraphics[scale=0.8]{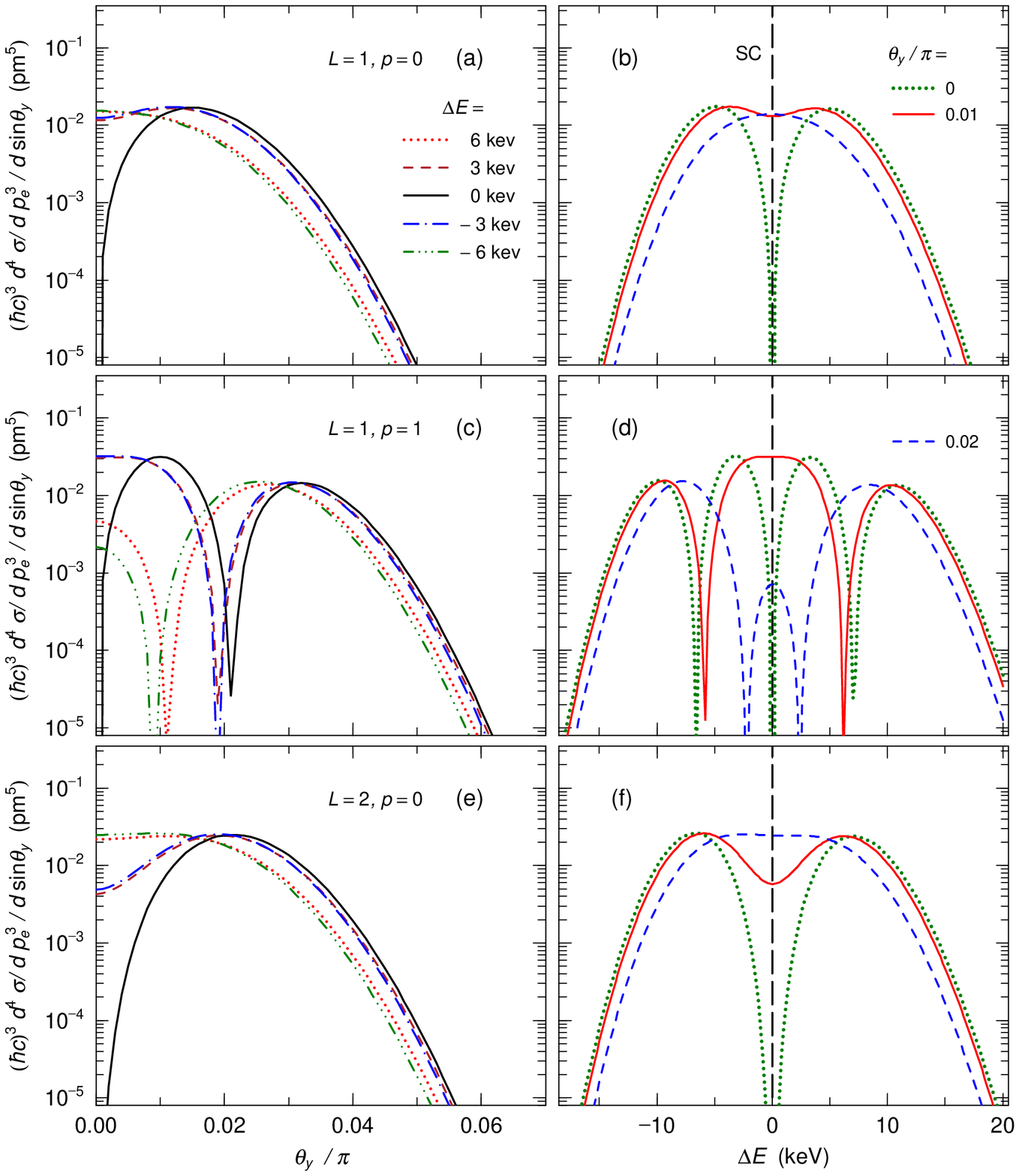}}
\caption{\small
The differential cross-section of Compton scattering
when  $\cos \theta_e = 0.95$.
The left panels show the $\theta_y$  dependence
when $L=1$, $p = 0$ (a), $L=1$, $p = 1$ (c), and $L=2$, $p = 0$.
The dotted, dashed, solid, dot-dashed, and dpt-dot-dashed lines represent the results when $\Delta E = 10$~keV, 
5~keV, 0~keV, $-5$~keV, and $-10$~keV, respectively.
The right panels show
its energy spectra of the scattered photons,
when $L=1$, $p = 0$ (b), $L=1$, $p = 1$ (d), and
 $L=2$, $p = 0$  (f).
The dotted, solid, and dashed lines represent 
the results when $\theta_y / \pi = 0$, 0.01, and 0.02, respectively.
The long dashed lines indicate the results in the standard 
Compton (SC) scattering.
}
\label{d4SigAE}
\end{center}
\end{figure}

Let us discuss the result quantitatively.
We show the differential cross-sections at $\cos \theta_e = 0.95$ when
$L=1$ and $p=0$ (a,b), $L=1$ and $p=1$ (c,d) and $L=2$ and $p=0$ (e,f)
in Fig.~\ref{d4SigAE}.  
The left panels (a,c,e) present the angular dependence of
the scattered photons.
It is again confirmed the result that the strengths are exactly zero at 
$\phi_y = 0$ and $\Delta E = 0$ (see the solid lines), at which the strength appears 
in the standard Compton scattering.
The position of the peak depends on $L$.
In the case of (a), the peak with $\Delta E=0$ is located at $\phi_y$/$\pi = 0.015$.
As $L$ increases, 
the peak position shifts toward a larger polar angle 
[see Figs.~\ref{d4SigAE}(e)],
which also corresponds to the shape of the photon wave function
represented by the Laguerre function $\el^{L}_p$.
Furthermore, the number of the peaks correlates with the node, $p$.
In the case of (c), there are two peaks, which originates from the fact that
the photon wave function has a node in the transverse direction for $p=1$.
%
%As the $\theta_e$ increases, furthermore, the spread of the $\theta_y$
%dependence becomes smaller (see (b), (d) and (f)). 
%The figures for $\cos \theta_e = 0.95$ show the similar trend 
%[see (b), (d) and (f)].
 %
These results indicate that, by measuring the angles of the scattered
photon and electron, 
one can identify the angular momentum and the node of the LG wave 
for incident photon. 

The right panels (b,d,f) present the expected energy spectra of 
the scattered photons.
In the case of the standard Compton scattering, the energy is 
uniquely determined (see long dashed-lines); 
however, the energy of $\gamma$-ray vortices spreads.
This result indicates another method to identify $\gamma$-ray vortex in
the measurements: the energy and angle of 
the scattered photon coincidentally with the angle of the scattered electron.
%
%Here, we give a comment.
Finally, we point out that the cross-sections are dependent on the scattered electron angle,
thgouh we show only the results at $\cos \theta_e = 0.95$.
As $\theta_e$ decreases, the energy distributions become broader while the $\theta_y$ distributions become narrower.

The $\gamma$-ray vortex is expected to be generated by laboratory
experiments in the near future.
In the present study, we study the Compton scattering 
with the gamma-ray vortex with the LG wave function, 
thereby find two unexpected features.
First, the differential cross-sections have finite values out of
the $zx$-plane.
Second, the energy of the scattered photon shifts from
that of the standard Compton scattering.
These results indicate that Compton scattering with $\gamma$-ray
vortices is useful to identify the nature of the photon vortices,
the node $p$ and the angular momentum $L$, 
when the final electron and photon are coincidentally measured. 

\bigskip

%\noindent
%{\bf Acknowledgements}\\
This work was supported by Grants-in-Aid 
for Scientific Research of JSPS ( 15H03665, 16K05360, 17K05459.).

\begin{thebibliography}{99}

\bibitem{Allen92} L. Allen, {\it et al}., Phys. Rev. A {\bf 45}, 8185 (1992).

\bibitem{BPA94} 
M. Babiker, W.~L. Power, and L. Allen, Phys. Rev. Lett. {\bf 73}, 1239 (1994).

\bibitem{ACD06} 
A. Alexandrescu, D. Cojoc, and E.~I. DiFabrizio, Phys. Rev. Lett. {\bf 96}, 243001 (2006).

\bibitem{PBMVPCR10} 
A. Picon, {\it et al}.,  New J. Phys. {\bf 12}, 083053 (2010).

\bibitem{Yao11}
A.~M. Yao and M.~J. Padgett, Adv. Opt. Photon. {\bf 3}, 161 (2011).

\bibitem{CWSJZ12} 
X. Cai, {\it et al.},  Science. {\bf 338}, 363 (2012).

\bibitem{Wang12} 
J. Wang, {\it et al.}, Nat. Phot. {\bf 6}, 488 (2012).

\bibitem{LSBP13}
M.~P.~J. Lavery,  {\it et al.}, Science {\bf 341}, 537 (2013).

\bibitem{TTTTMMO13} 
K. Toyoda, {\it et al.},  Phys. Rev. Lett. {\bf 110}, 143603 (2013).

\bibitem{Afanasev13}
A. Afanasev, C.~E, Carlson, and A. Mukherjee, Phys. Rev. A. {\bf 88} 033841 (2013).

\bibitem{BYRTKHWR13} 
N. Bozinovic, {\it et al.}, Science {\bf 340}, 1545 (2013).

\bibitem{TTMA11}
F. Tamburini, Nat. Phys. {\bf 7}, 195 (2011).

\bibitem{Jentschura11a}
U.~D. Jentschura and V.~G. Serbo,  Phys. Rev. Lett. {\bf 106}, 013001 (2011).

\bibitem{Jentschura11b}
U.~D. Jentschura and V.~G. Serbo, Eur. Phys. J. C {\bf 71}, 1571 (2011).

\bibitem{Petrillo16}
V. Petrillo, {\it et al.},  Phys. Rev. Lett. {\bf 117}, 123903 (2016).

\bibitem{Taira17}
Y. Taira, T. Hayakawa, and M. Katoh, Sci. Rep. {\bf 7}, 5018 (2017).

\bibitem{Ivanov11}
I.~P. Ivanov,  Phys. Rev. D {\bf 83}, 093001 (2011).

\bibitem{Jones95}
P.~M. Jones,  {\it et al.},  Nucl. Instr. Methods Phys. Res. A {\bf 362}, 556 (1995).

\bibitem{Lei97}
F. Lei, A.~J. Deam, and G.~L. Hills, Spac. Sci. Rev. {\bf 82}, 309 (1997).

\bibitem{Bjorken}
S.~D. Bjorken and S.~D.~Drell, "Relativistic Quantum Fields (Dover Books on Physics)", Dover Publications (November 25, 2014)

\end{thebibliography}
\end{document}